\pdfoutput=1

\documentclass[%
reprint,
showpacs,preprintnumbers,
showkeys,
amsmath,amssymb,
aps,
prl,
floatfix,
longbibliography,
]{revtex4-2}

\usepackage[english]{babel}
\usepackage[utf8]{inputenc}
\usepackage{graphicx}
\usepackage{bm}
\usepackage{xcolor}
\usepackage{epsfig}
\usepackage{amsmath}
\usepackage{amsfonts}
\usepackage{amssymb}
\usepackage{amsthm}
\usepackage{MnSymbol}
\usepackage{hyperref}
\usepackage{tikz}
\usepackage{dsfont}
\usepackage[all]{nowidow}
\usepackage{pifont}
\usepackage{algorithm}
\usepackage{algpseudocode}
\usepackage[capitalise]{cleveref}
\usepackage[export]{adjustbox}
\usepackage{xspace}
\usepackage{nicefrac}

\usepackage{doi}
\usepackage{lipsum}

\crefname{section}{Supplement}{Supplements}%
\Crefname{section}{Supplement}{Supplements}%
\crefname{subsection}{Supplement}{Supplements}%
\Crefname{subsection}{Supplement}{Supplements}%
\crefname{subsubsection}{Supplement}{Supplements}%
\Crefname{subsubsection}{Supplement}{Supplements}%

\usepackage[letterspace=130]{microtype}

\definecolor{refColor}{HTML}{0376E9}
\definecolor{figColor}{HTML}{E90303}
\definecolor{urlColor}{HTML}{0bcb9a}

\hypersetup{
    unicode=false,                 
    pdftoolbar=true,               
    pdfmenubar=true,               
    pdffitwindow=false,            
    pdfstartview={FitH},           
    pdftitle={},                   
    pdfauthor={Nicolai Lang},      
    pdfsubject={Quantum physics},  
    pdfcreator={Nicolai Lang},     
    pdfproducer={Nicolai Lang},    
    pdfkeywords={},                
    pdfnewwindow=true,             
    colorlinks=true,               
    linkcolor=figColor,            
    citecolor=refColor,            
    filecolor=magenta,             
    urlcolor=urlColor              
}

\newcommand{\bra}[1]{\mathinner{\langle{#1}|}}
\newcommand{\ket}[1]{\mathinner{|{#1}\rangle}}

\renewcommand{\vec}[1]{\mathbf{#1}}

\renewcommand{\vec}[1]{\boldsymbol{#1}}

\newcommand{\tr}[1]{\operatorname{Tr}\left[#1\right]}

\newcommand{\Tr}[2]{\operatorname{Tr}_{#2}\left[#1\right]}

\renewcommand{\S}{\mathcal{S}}
\newcommand{\D}{\mathcal{D}}
\newcommand{\R}{\mathcal{R}}

\newcommand{\CaptionMark}[1]{\textit{#1}}

\graphicspath{{./fig/}{./data/}}

\newcommand{\papertitle}{Robust detection of an entanglement transition in the \\ projective transverse field Ising model}
\newcommand{\paperauthors}{%
\author{Felix Roser}
\email{felix.roser@itp3.uni-stuttgart.de}
\author{Etienne M. Springer}
\author{Hans Peter Büchler}
\author{Nicolai Lang}
}
\newcommand{\paperaffiliations}{%
\affiliation{%
	Institute for Theoretical Physics III 
	and Center for Integrated Quantum Science and Technology,\\
	University of Stuttgart, 70550 Stuttgart, Germany
	}
}

\begin{document}

\begin{abstract}
	
We propose a scalable and noise-resilient protocol for the detection of the
entanglement transition in a projective version of the transverse field Ising
model. Entanglement transitions are experimentally difficult to observe due to
the inherent randomness of projective measurements and noise in large-scale experimental settings. Our
approach combines error correction algorithms with classical shadow
tomography to overcome both problems. This allows for experimentally accessible
upper and lower bounds on the entanglement transition without postselection or
full state tomography. These bounds remain robust under noise and their
sharpness is a measure of the noise rate.

\end{abstract}

\title{\papertitle}
\paperauthors
\paperaffiliations
\date{\today}
\maketitle

\paragraph{Introduction.}

The creation of entanglement is typically associated with multi-qubit
unitaries, whereas projective single-qubit measurements cause its destruction.
Remarkably, the study of \emph{random quantum circuits} revealed that this
competition can drive non-equilibrium phase transitions between phases with
volume-law and area-law entanglement \cite{Li_2018, Li_2019, Skinner_2019,
Potter_2022, Fisher_2023}. Similar transitions can also occur between different
area-law phases, and be driven by the competition of projective measurements
only \cite{Lang_2020, Nahum_2020, Ippoliti_2021_EPT}. 
Entanglement transitions are characterized by changes in the scaling of
pure-state entanglement entropies averaged over many random quantum circuits
\cite{Neumann_1932, Bennett_1996, Li_2019, Lang_2020} -- while there is no
signature in the density matrix of the ensemble of trajectories itself. This
makes entanglement transitions experimentally elusive
\cite{Noel_2022,Koh_2023,Hoke_2023}:
Measuring entanglement requires multiple copies of a quantum state, whereas
preparing such copies with random quantum circuits is exponentially unlikely
due to the inherent randomness of measurements. This \emph{postselection
problem} bars straightforward observations of entanglement transitions
\cite{Czischek_2021,Koh_2023}. 
In this paper, we address this issue by combining shadow tomography
\cite{Garratt_2024} with tailored error correction algorithms \cite{Roser_2023}
to construct protocols for probing a particular entanglement transition
reliably in large, noisy systems.

Various strategies have been suggested to either alleviate or bypass the
postselection problem, each with benefits and downsides.
This includes unitary ``stand-ins'' that map to projective time evolutions
\cite{Ippoliti_2021, Ippoliti_2022, Hoke_2023}, quantum steering via unitary
feedback \cite{Roy_2020, Noel_2022, Pöyhönen_2025}, signatures in more
accessible quantities like cross entropies \cite{Li_2023, Tikhanovskaya_2024,
Kamakari_2025}, studying special non-local models \cite{Passarelli_2024,
Feng_2023, Feng_2025}, and exploiting emergent quantum error correction
abilities \cite{Gullans_2020, Gullans_2020_2, Roser_2023,Ippoliti_2024,
Dehghani_2023, Khindanov_2025}. Following a different route, shadow tomography
\cite{Aaronson_2020, Huang_2020, Elben_2023} has been proposed as a tool to
measure upper and lower bounds on entanglement entropies \cite{Garratt_2024,
McGinley_2024}.
Despite this progress, experimentally feasible protocols for probing
entanglement transitions that are both scalable and noise-resilient are still
lacking.

In this paper, we combine efficient error correction algorithms
\cite{Roser_2023} with classical shadow tomography \cite{Garratt_2024} to
contrive explicit, platform-agnostic, robust protocols for measuring upper and
lower bounds on a particular entanglement transition:
We consider the paradigmatic \emph{projective transverse field Ising
model}~\cite{Lang_2020,Nahum_2020} which features a transition between two
area-law phases and is both experimentally accessible and efficiently
simulatable. We show how decoding quantum information can be used to measure a
\emph{lower bound} on the entanglement transition, and then contrive a protocol
based on error correction and shadow tomography \cite{Garratt_2024} to measure
an \emph{upper bound} as well. In combination, both procedures yield an
experimentally accessible, robust \emph{interval} that constrains the
entanglement transition and probes the noise rate.

\begin{figure}[tbp]
	\centering
	\includegraphics[width=0.95\linewidth]{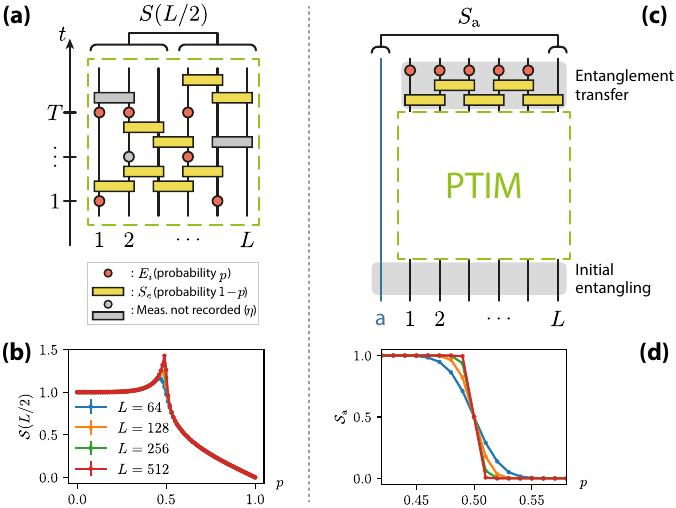}
	\caption{%
		\CaptionMark{Entanglement transition of the PTIM.}
		(a)~Exemplary trajectory of a PTIM. Circles (boxes) correspond to
		projective $E_i\sim X$ ($S_e\sim ZZ$) measurements applied with probability
		$p$ ($1-p$) per site and time step. Noise manifests as unrecorded
		measurements (gray circles and boxes).
		(b)~The PTIM features a transition in the trajectory-averaged half-system
		EE $\mathcal{S}(L/2)$ in the steady state for $T\to\infty$ (here $T\gtrsim
		10L$) at $p_c=0.5$ between two area-law phases.
		(c)~Alternative protocol to detect the transition utilizing an intially
		fully-entangled ancilla qubit.
		(d)~The transition in the survival probability of an initial Bell cluster
		is measured by the EE $\mathcal{S}_\text{a}$ of the ancilla for $L=T$.
		All simulations in this paper are based on $10^5$ trajectories. 
	}
	\label{fig:PTIM}
\end{figure}

\paragraph{Setting.}

The projective transverse field Ising model (PTIM) \cite{Lang_2020,Nahum_2020}
is defined on an open chain of $L$ qubits that are initialized in a state
$\ket{\Psi_0}$. Let $X_i$ and $Z_i$ denote Pauli matrices acting on qubit
$i\in\left\{1,\ldots,L\right\}$. A \emph{trajectory} $\ket{\Psi_t}$ of the PTIM
is generated by a discrete time evolution: Let the system be in state
$\ket{\Psi_t}$ at time step $t$. The next state $\ket{\Psi_{t+1}}$ is obtained
by first looping over all sites and measuring with probability
$p\in\left[0,1\right]$ the operators $E_i\equiv X_i$ projectively, and then
looping over all pairs of adjacent sites $e=(i,i+1)$ (edges) and measuring
$S_e\equiv Z_iZ_{i+1}$ with probability $1-p$. This is repeated for $T$ time
steps to produce the final state $\ket{\Psi_T}$. The measurements of a
trajectory are recorded, where $M^\text{p}\equiv(E^\text{p},S^\text{p})$ denotes the
measurement \emph{pattern} (locations and times of measurements) and
$M^\text{r}\equiv(E^\text{r},S^\text{r})$ contains their \emph{results}.
In an experiment, we expect the sampling of a trajectory to be error prone.
Here we assume that errors manifest as unrecorded measurements: With
probability $\eta\in [0,1]$ (noise rate), projective measurements $E_i$ and
$S_e$ that were applied are \emph{not recorded} in the history
$M^{\text{p},\text{r}}$. Hence the generated ensemble of trajectories depends
on the initial state $\ket{\Psi_0}$, the spacetime dimensions $L$ and $T$, and
the measurement probability $p$, while the noise rate $\eta$ only affects the
measurement record $M^{\text{p},\text{r}}$. A \emph{sample}
$s\equiv(\ket{\Psi_T},M^\text{p,r})$ is then given by the final state and the
measurement record. An exemplary trajectory of a noisy PTIM is shown in
\cref{fig:PTIM}~(a).

Consider a subsystem $A\subseteq\{1,\ldots,L\}$ of $\lvert A\rvert$ qubits. For
each sample $s$, we define the reduced density matrix
$\rho_{s,A}:=\operatorname{Tr}_{\overline{A}}\left[\ket{\Psi_T}\bra{\Psi_T}\right]$
and the von Neumann entanglement entropy (EE)
$S_s(A):=-\tr{\rho_{s,A}\log_2\rho_{s,A}}$. We are interested in the average of
entanglement entropies $\S(A)=\llangle S_s(A)\rrangle$ over many ($\sim 10^5$)
samples. In the following, quantities of single samples are labeled by an index
$s$ while sample averages are denoted by $\llangle\cdot\rrangle$ and
calligraphic symbols.
The entanglement transition of the PTIM manifests in the singular behavior of
the half-system EE $\S(L/2)\equiv\S(\{1,\ldots,L/2\})$ as a function of the
measurement probability $p$ in the steady state limit $T\to\infty$ of large
systems, see \cref{fig:PTIM}~(b) for simulations initialized in a global GHZ
state $\ket{\Psi_0}\propto\ket{0\ldots0}+\ket{1\ldots1}$. The critical point
$p_c=0.5$ corresponds to a bond-percolation transition and is described by a
conformal field theory \cite{Lang_2020,Nahum_2020}; it separates two area-law
phases with ($p<p_c$) and without ($p>p_c$) long-range entanglement.

PTIM trajectories can be simulated efficiently via the stabilizer formalism
\cite{Gottesman_1996, Gottesman_1998, Gottesman_1998_Theory, Aaronson_2004, Nielsen_2010};
however, since we are only interested in the entanglement of states, simulating
a \emph{colored cluster model} (CCM) is even more efficient \cite{Lang_2020}.
This method assumes initial states of the form
$\ket{\Psi_0}\propto\ket{0\ldots0}\pm\ket{1\ldots1}$ and exploits that the PTIM
only produces product states of \emph{Bell clusters} $\ket{\vec
m}_A\pm\ket{\overline{\vec m}}_A$, with $\vec m$ the configuration of some
subset $A$ of qubits and $\overline{\vec m}$ its bitwise complement. Indeed,
$E_i$-measurements ``poke holes'' into Bell clusters by projecting qubit $i$
into a single-site Bell cluster $\ket{\pm}_i\propto\ket{0}_i\pm\ket{1}_i$
whereas $S_e$-measurements merge adjacent Bell clusters. Since entanglement is
independent of the configurations $\vec m$ and relative signs $\pm$, the
entanglement structure can be tracked by ``coloring'' qubits according to the
Bell cluster to which they belong at any given time, and evolving this pattern
by updating colors according to the described effects of measurements. In the
entangling phase of the PTIM ($p<p_c$) the survival probability of the initial
Bell cluster $\ket{\Psi_0}\propto\ket{0\ldots0}\pm\ket{1\ldots1}$ approaches
one for $L=T\to\infty$, whereas in the disentangling phase ($p>p_c$) it goes to
zero.

This motivates an alternative probe of the entanglement transition,
\cref{fig:PTIM}~(c). A single \emph{ancilla qubit} $i=\text{a}$ is added to the
$L$ \emph{system qubits} and included in the initial Bell cluster
$\ket{\Psi_0}\propto\ket{0}_\text{a}\ket{0\ldots0}_L+\ket{1}_\text{a}\ket{1\ldots1}_L$
by an initial entangling step. After this initialization (which we assume to be
perfect), the ancilla remains unaffected by the PTIM evolution. After $T=L$
time steps, the trajectory is finalized by an (also perfect) \emph{entanglement
transfer}, where first all $S_e$-measurements on system qubits are applied
(merging all Bell clusters), followed by all but the last $E_i$-measurements
($i\neq \text{a},L$). We record the outcomes of these additional measurements
in $M^\text{r}=(E^\text{r},S^\text{r})$ for later use. As a final step, the EE
$S_s(\{\text{a}\})=S_s(\{L\})$ of the ancilla is evaluated.
The rationale is that the initial Bell cluster survives the PTIM evolution
\emph{if and only if} the ancilla remains entangled with the system; i.e., the
sample average $\S_\text{a}\equiv\llangle S_s(\{\text{a}\})\rrangle$ directly
probes the survival probability of the initial Bell cluster. Note that the
entanglement transfer step is not needed for the measurement of $\S_\text{a}$
but becomes important later for determining bounds on the entanglement
transition via shadow tomography.
We show simulations of $\S_\text{a}$ in \cref{fig:PTIM}~(d). In contrast to the
half-system EE $\S(L/2)$, the ancilla EE features a crossing for different
system sizes at $p_c=0.5$, clearly indicating the transition. Note that
$\S_\text{a}$ only requires the measurement of the EE of a \emph{single} qubit.

\paragraph{Postselection problem.}

Numerically, the PTIM is easily accessible. However, to verify the predictions
in \cref{fig:PTIM}~(b) or (d) \emph{experimentally}, one must measure
entanglement entropies of each final state $\ket{\Psi_T}$. This is notoriously
difficult, as it in general requires full state tomography which, in turn, is
only accessible through many copies of states $\ket{\Psi_T}$ \cite{ODonnell_2016, Scharnhorst_2025}.
Unfortunately, these states are produced by a process that is inherently
probabilistic for two reasons. The measurement \emph{pattern} $M^p$ is random,
but this randomness is (for perfect measurements) assumed to be under the
experimentalists control. By contrast, the measurement \emph{results} $M^r$
are random according to the Born rule and cannot be controlled \emph{in
principle}.
Since final states $\ket{\Psi_T}$ depend on these measurement outcomes, it is
exponentially unlikely (in $T\sim L$) to prepare multiple copies -- which
renders entanglement entropies experimentally inaccessible.
In the remainder of this paper, we show how to alleviate this
\emph{postselection problem} and make the entanglement transition of the PTIM
experimentally accessible, even for unreliable measurement records $M^{p,r}$.

\begin{figure}[tbp]
	\centering
	\includegraphics[width=0.9\linewidth]{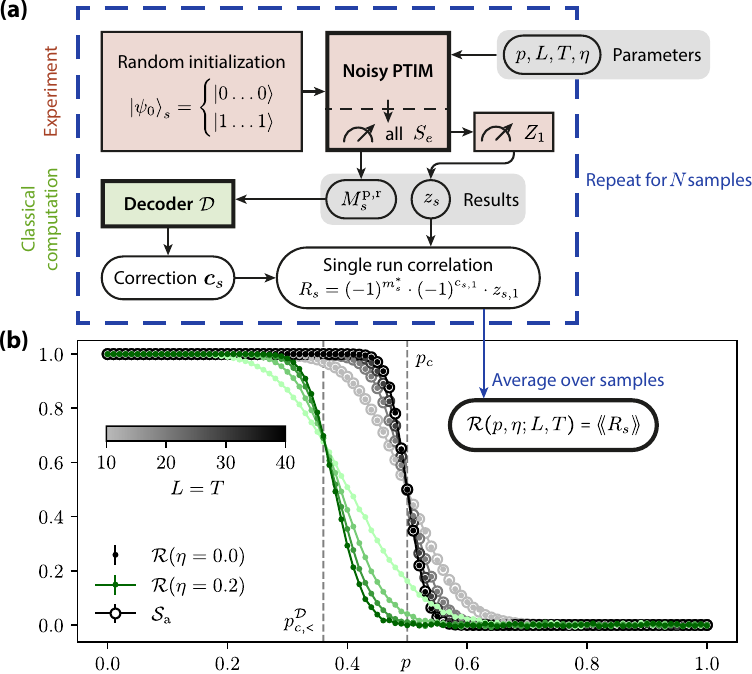}
	\caption{%
		\CaptionMark{Lower bound via decoding.}
		(a)~Protocol to measure the decoding transition of the PTIM as described in
		the text.
		(b)~Sample-averaged correlation as a function of $p$ for system sizes
		$L=10,\ldots,40$ with (green) and without (black) noise. In a noisy
		circuit, the transition is shifted to the left and provides a lower bound
		on the entanglement transition.
	}
	\label{fig:decoding}
\end{figure}

\paragraph{Lower bound via decoding.}
\label{sec:InformationRetrievalAsLowerBound}

It is known that entangling phases of random quantum circuits behave as
emergent quantum codes \cite{Gullans_2020, Gullans_2020_2}, i.e., amplitudes
suitably encoded in the initial state survive for exponentially long times --
despite finite rates of projective measurements!
Indeed, the initial state of the PTIM can encode a single logical qubit via
$\ket{\Psi_0}=\alpha\ket{0\ldots 0}+\beta\ket{1\ldots 1}$ with amplitudes
$|\alpha|^2+|\beta|^2=1$. As shown in Ref.~\cite{Lang_2020}, these amplitudes
survive the projective dynamics if an initial Bell cluster would, that is, in
the entangling phase ($p<p_c$). In this case, and if the trajectory is
finalized by one round of deterministic $S_e$-measurements to merge all Bell
clusters, the final state has the form
$\ket{\Psi_T}=\alpha\ket{\vec{m}}+\beta\ket{\overline{\vec{m}}}$, where the
basis configuration $\vec{m}\in\{0,1\}^L$ depends on the measurement pattern
and results of the trajectory. By contrast, in the disentangling phase
($p>p_c$), the probability for the amplitudes to be destroyed approaches one
for $L=T\to\infty$ and the final state has the form
$\ket{\Psi_T}=\ket{\vec{m}}\pm\ket{\overline{\vec{m}}}$. 
This suggests to \emph{certify} the entangling phase by successfully retrieving
information from the final state $\ket{\Psi_T}$ that was encoded in the initial
state $\ket{\Psi_0}$. Since this decoding can fail even when the amplitudes
survive, such a protocol provides in general an experimentally accessible
\emph{lower-bound} (in $p$) on the entanglement transition. 

We now describe such a protocol for the PTIM, \cref{fig:decoding}~(a).
For each sample $s$, we randomly encode one bit $m_s^*\in\{0,1\}$ into the
initial state $\ket{\Psi_0}_s=\ket{m_s^*,\ldots,m_s^*}$ \footnote{%
	Note that $X_1X_2 \cdots X_L$ is a symmetry of the PTIM dynamics; hence an
	encoding in symmetry eigenstates $\ket{\Psi_0}\propto \ket{0\ldots
	0}+(-1)^m\ket{1\ldots 1}$ cannot be used to detect the entanglement
	transition.
}, 
and then perform a PTIM evolution for $T=L$ steps. Each trajectory is finalized
by a perfect measurement of all $S_e$-operators, so that
$\ket{\Psi_T}_s=\ket{\vec m_s}$ with a random bit pattern $\vec m_s$ if the
amplitudes survive, and $\ket{\Psi_T}_s=\ket{\vec m_s}\pm\ket{\overline{\vec m}_s}$
otherwise. Finally, we measure $Z_1$ on the first qubit. Thus, for each sample,
we possess the following data: (1) the encoded bit $m_s^*$, (2) the (potentially
incomplete) measurement record $M_s^{\text{p},\text{r}}$, and (3), the outcome
$z_{s,1}\in\{\pm 1\}$ of the final $Z_1$-measurement.
Our task is to recover $m_s^*$ from $M_s^{\text{p},\text{r}}$ and $z_{s,1}$
(referred to as \emph{decoding}). If decoding succeeds with high probability,
the system must be in the entangling phase. 
A \emph{decoder} is an algorithm $\D(M_s^{\text{p},\text{r}})=\vec c_s$ that
uses the measurement record to compute a \emph{correction} $\vec c_s$ such that
$\vec c_s\oplus\vec m_s=(m_s^*,\ldots,m_s^*)$ with high probability (if the
amplitudes survive). Here, $\oplus$ denotes bit-wise addition modulo-two. We
can quantify the success of decoding by evaluating the correlation
\begin{align}
	R_s:=(-1)^{m_s^*}\cdot\left(-1\right)^{c_{s,1}}\cdot z_{s,1}\,,
\end{align}
averaged over many samples: $\mathcal{R}(p,\eta;L,T)\equiv\llangle
R_s\rrangle$. For perfect measurement records ($\eta=0$), decoding is trivial
because the trajectory can be simulated exactly and efficiently via the
stabilizer formalism [or an even more efficient \emph{extended colored cluster
simulation} (ECCS), see \cref{subsubsec:ModifiedColoredClusterModel}].
In \cref{fig:decoding}~(b) we plot $\mathcal{R}$ for this case. As expected,
the (accessible!) decoding transition of $\mathcal{R}$ perfectly matches the
(inaccessible) entanglement transition of $\S_\text{a}$.

\begin{figure*}[tbp]
	\centering
	\includegraphics[width=1.0\linewidth]{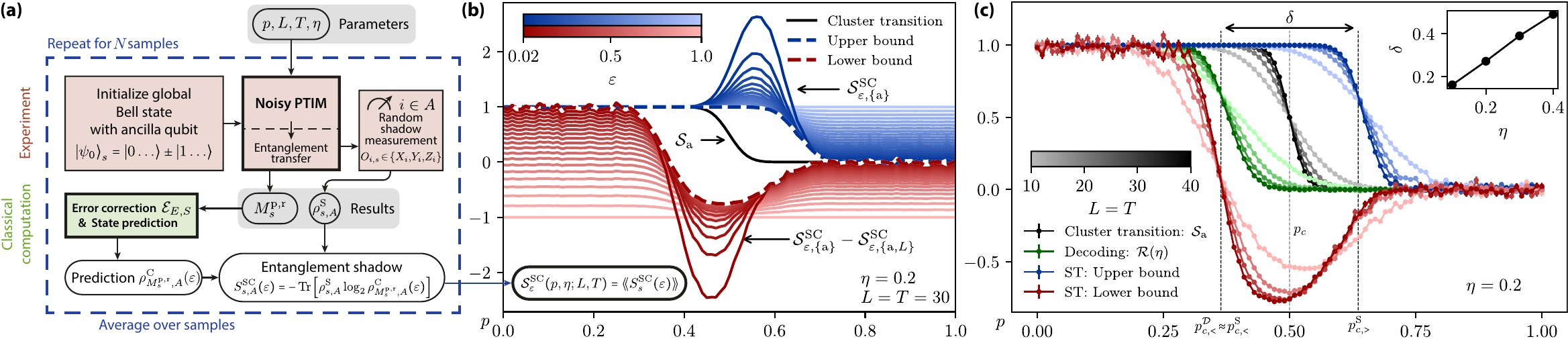}
	\caption{%
		\CaptionMark{Upper and lower bound via shadow tomography.}
		(a)~Protocol to measure the shadow of the entanglement entropy of a
		subsystem $A$ on the noisy PTIM, based on a decoding algorithm for
		classical state prediction (see text).
		(b)~Optimization over the regularization $0<\varepsilon<1$ yields an upper
		(dashed blue) and lower bound (dashed red) on the ancilla EE $\S_\text{a}$
		(solid black), shown for a finite noise rate $\eta=0.2$ and $L=T=30$.
		%
		(c)~Comparison of the lower bounds from decoding (green) and ST (red), and
		the upper bound from ST (blue). For $L=T=10,\ldots,40$ we find clear
		crossings which provide experimentally accessible lower and upper bounds on
		the entanglement transition at $p_c=0.5$ (black). The simulations suggest
		that the two lower bounds share the same threshold. The interval $\delta$
		between upper and lower bounds can be used to measure the noise rate $\eta$
		(inset).
	}
	\label{fig:shadow}
\end{figure*}

\paragraph{Decoding with noise.}

If the measurement record $M_s^{\text{p},\text{r}}$ is \emph{incomplete}
($\eta\neq 0$), a straightforward simulation is impossible since the results
$M_s^\text{r}$ can be inconsistent with the recorded pattern $M_s^\text{p}$. To
fix this, we extend the decoder $\D$ (used for the noise-free case) by an
additional preprocessing step $\mathcal{E}$ which augments the
observed history $M_s^{\text{p},\text{r}}$ to a hypothetical history $\tilde
M_s^{\text{p},\text{r}}$ that is both \emph{likely} and \emph{consistent}; the
computation of the correction $\vec c_s$ (again via an ECCS) is then based on
this consistent history. We denote this error-corrected decoder by $\D_E$.
Since the \emph{outcomes} of hypothetical measurements cannot be predicted
(this follows from Pauli measurements either commuting or anticommuting) and
$\vec c_s$ is computed from the results of $S_e$-measurements, it is sufficient
to augment $E_s^\text{p}\mapsto\tilde E_s^\text{p}$ by hypothetical
$E_i$-measurements (without results). To construct a consistent pattern $\tilde
M_s^{\text{p}}\equiv (\tilde E_s^\text{p},S_s^\text{p})$ with as few unobserved
$E_i$-measurements as possible, the error correction algorithm
$\mathcal{E}\equiv\mathcal{E}_E$ employs \emph{minimum-weight perfect matching}
(MWPM)~\cite{Edmonds_1965, Kolmogorov_2009}, see \cref{app:DecodingErrors}. 

Using this decoder for the protocol in \cref{fig:decoding}~(a), we can again
measure $\R$; results for $\eta=0.2$ are shown in \cref{fig:decoding}~(b).
The decoding threshold $p_{c,<}^{\mathcal{D}}$ of the MWPM-based decoder $\D_E$
is clearly indicated by a crossing of $\R(p)$ for $L=T\to\infty$. Since a
necessary (but not sufficient) condition for successful decoding is the
survival of an initial Bell cluster, it is $\R(p)\le \S_\text{a}(p)$ in the
limit of many samples, with equality only for $\eta=0$. Hence
$p_{c,<}^{\mathcal{D}}\leq p_c$ provides a robust, experimentally accessible
lower-bound on the entanglement transition. The deviation
$|p_{c,<}^{\mathcal{D}}-p_c|$ depends both on the noise $\eta$ and the decoding
algorithm. Since $\D_E$ is not necessarily a maximum likelihood decoder,
smaller deviations might be achievable (though significant improvements are
unlikely \cite{Roser_2023,Bravyi_2014}).

\paragraph{Upper bound via shadow tomography.}

For an experimentally accessible \emph{upper} bound, we employ the shadow
tomography (ST) protocol proposed by Garratt and Altman \cite{Garratt_2024} and
combine it with efficient MWPM-based error correction for resilience against
noise.
We consider the setting depicted in \cref{fig:PTIM}~(b) with an initially
entangled ancilla qubit. The goal is to reliably bound its entanglement
entropy $\S_\text{a}$ from above by an experimentally accessible quantity which
can be obtained by the protocol in \cref{fig:shadow}~(a).
For every trajectory, the system is initialized in a fully entangled Bell state
and evolved under a PTIM circuit for $L=T$ time steps. A final entanglement
transfer [\cref{fig:PTIM}~(b)] ensures that the entanglement entropy
$S_\text{a}$ of the ancilla with the system qubit $i=L$ encodes the survival
of the initial Bell cluster; we denote the final state of these two qubits as
$\ket{\Psi_T^{\{\text{a},L\}}}_s$.
Due to the post selection problem, $S_\text{a}$ is not experimentally
accessible. Instead, we evaluate two auxiliary quantities that combined yield
an upper bound on $\S_\text{a}=\llangle S_\text{a}\rrangle$:

First, we randomly pick
$O_{\text{a},s}\in\left\{X_\text{a},Y_\text{a},Z_\text{a}\right\}$ and measure
this operator on the ancilla in the final state with outcome $o_s\in\{\pm 1\}$;
from this, we compute the \emph{shadow}
$\rho_{s,\left\{\text{a}\right\}}^\text{S}:=\tfrac{3}{2}\left(\mathds{1}+o_s\cdot
O_{\text{a},s}\right)-\mathds{1}$ \cite{Huang_2020}.
Second, we use the history $M_s^{\text{p},\text{r}}$ to compute a
\emph{classical prediction}
$\rho_{M_s^{\text{p},\text{r}},\left\{\text{a}\right\}}^\text{C}$ for the state
$\rho_{s,\left\{\text{a}\right\}}\equiv\Tr{\ket{\Psi_T^{\{\text{a},L\}}}\bra{\Psi_T^{\{\text{a},L\}}}_s}{\{L\}}$
of the ancilla. We describe the algorithm for this prediction below.
Finally, the \emph{shadow of the ancilla EE} is defined as
\begin{align}
	S_{s,\left\{\text{a}\right\}}^\text{SC}
	:=-\tr{\rho_{s,\left\{\text{a}\right\}}^\text{S}
	\log_2\rho_{M_s^{\text{p},\text{r}},\left\{\text{a}\right\}}^\text{C}}
	\,.
	\label{eq:EntanglementShadow}
\end{align}
Note that for pure
$\rho_{M_s^{\text{p},\text{r}},\left\{\text{a}\right\}}^\text{C}$ this quantity
diverges, which is avoided by regularizing the prediction
$\rho_{M_s^{\text{p},\text{r}},\left\{\text{a}\right\}}^\text{C}(\varepsilon)
:=(1-\varepsilon)\rho_{M_s^{\text{p},\text{r}},\left\{\text{a}\right\}}^\text{C}+\tfrac{\varepsilon}{2}\mathds{1}$
via a depolarizing channel of strength $\varepsilon$. The regularized
shadow~\eqref{eq:EntanglementShadow} is then averaged over trajectories to
evaluate
$\S_{\varepsilon,\left\{\text{a}\right\}}^\text{SC}(p,\eta;L,T)\equiv\llangle
S_{s,\left\{\text{a}\right\}}^\text{SC}(\varepsilon)\rrangle$.
Crucially, in the limit of many samples, this quantity upper bounds the ancilla
entropy \cite{Garratt_2024}
\begin{align}
	\S_\text{a}
	\leq
	\S_{\varepsilon,\left\{\text{a}\right\}}^\text{SC}
	\,,
	\label{eq:ShadowUpperBound}
\end{align}
independent of $\varepsilon$ and the algorithm used for
$\rho_{M_s^{\text{p},\text{r}},\left\{\text{a}\right\}}^\text{C}$.
This follows from the properties of the shadow
$\rho_{s,\left\{\text{a}\right\}}^\text{S}$ and Klein's inequality \cite{Nielsen_2010}. 
Equality in \eqref{eq:ShadowUpperBound} is approached for ideal state
prediction,
$\rho_{M_s^{\text{p},\text{r}},\left\{\text{a}\right\}}^\text{C}\equiv\rho_{s,\left\{\text{a}\right\}}$,
in the limit $\varepsilon\to 0$ and infinitely many samples (possible only for
perfect measurement records, $\eta=0$).

The challenge is to make Garatt and Altman's protocol robust against noise
($\eta\neq 0$) for large systems ($L\to\infty$). To this end, we introduce an
efficient algorithm which combines classical simulations with error correction
(see \cref{app:ClassicalStatePredictionViaDecoding} for technical details):
To compute $\rho_{M_s^{\text{p},\text{r}},\left\{\text{a}\right\}}^\text{C}$,
we trace out the system qubit $i=L$ from a prediction for
$\ket{\Psi_T^{\{\text{a},L\}}}_s$. For this state, the PTIM evolution allows
\emph{eight} possible stabilizer groups, namely $\langle \pm Z_\text{a}Z_L,\pm
X_\text{a}X_L\rangle$ (entangled ancilla) or $\langle \pm X_\text{a},\pm
X_\text{a}X_L\rangle$ (disentangled ancilla). 
To decide on the group and the signs, we employ the error correction algorithm
$\mathcal{E}_E$ from above, and combine it with the ``dual'' algorithm
$\mathcal{E}_S$ which uses results of $E_i$-measurements to augment
$M_s^{\text{p},\text{r}}$ by unobserved $S_e$-measurements (again via MWPM,
\cref{app:DecodingStabilizers}).
If the algorithm concludes that the sign of a stabilizer cannot be derived,
the latter is dropped and a mixed state is predicted. This avoids false
predictions which lead to infinities for $\varepsilon\to 0$ (and thereby bad
upper bounds).

We plot $\S_{\varepsilon,\left\{\text{a}\right\}}^\text{SC}$ for $\eta=0.2$ in
\cref{fig:shadow}~(b) for different regularizations $0<\varepsilon\leq 1$.
Large $\varepsilon$ suppress statistical fluctuations, but lead to less sharp
bounds on the entanglement entropy if state predictions are reliable. Small
$\varepsilon$ \emph{can} yield tighter bounds, but false state predictions are
punished more severely [causing the peak in \cref{fig:shadow}~(b)].
Hence the optimal upper bound is the \emph{lower envelope} of the upper bounds
computed for different $0<\varepsilon\leq 1$. \cref{fig:shadow} (b) and (c)
demonstrate that this protocol provides a clear upper bound
$p_{c,>}^{\text{S}}$ (indicated by a crossing for $L=T\to\infty$) on the
entanglement transition at $p_c=0.5$. Notably, the protocol requires only a
single-site Pauli measurement $O_{\text{a},s}$ on the final state of each
sample. In \cref{app:ShadowTomographyWithoutDecoding} we demonstrate that
sophisticated error correction is indeed crucial for this result.

\paragraph{Alternative lower bound via shadow tomography.}

Shadow tomography can also be used to \emph{lower}-bound $\S_\text{a}$
\cite{Garratt_2024}. Here we compare this bound with our decoding bound
$p_{c,<}^{\mathcal{D}}$.
To compute the ST-based lower bound, one measures on
$\ket{\Psi_T^{\{\text{a},\text{L}\}}}_s$ also the system qubit $i=L$ in a
random basis to construct its shadow $\rho_{s,\left\{L\right\}}^\text{S}$. In
combination with the ancilla shadow, this yields the two-qubit shadow
$\rho_{s,\left\{\text{a},L\right\}}^\text{S}:=\rho_{s,\left\{\text{a}\right\}}^\text{S}\otimes\rho_{s,\left\{L\right\}}^\text{S}$.
The (regularized) state prediction
$\rho_{M_s^{\text{p},\text{r}},\left\{\text{a},L\right\}}^\text{C}(\varepsilon)$
for \emph{both} qubits is already provided by the error-corrected algorithm
sketched above. One can then generalize \cref{eq:EntanglementShadow} to the
shadow $S_{s,\left\{\text{a},L\right\}}^\text{SC}(\varepsilon)$ of the EE of
the \emph{combined system} $\{\text{a},L\}$, and finds for its sample average
the lower bound $\S_{\varepsilon,\left\{\text{a}\right\}}^\text{SC}-\S_{\varepsilon,\left\{\text{a},L\right\}}^\text{SC}\le \S_\text{a}$ \cite{Garratt_2024},
%
%
where $\S_{\varepsilon,\left\{\text{a}\right\}}^\text{SC}$ is the shadow used
for the upper bound.
Simulations for $\eta=0.2$ and $0<\varepsilon\leq 1$ are shown in
\cref{fig:shadow}~(b). Here, the \emph{upper envelope} yields the best lower
bound on $\S_\text{a}$. This envelope exhibits a clear crossing at
$p_{c,<}^{\text{S}}$ for $L=T\to\infty$, \cref{fig:shadow} (c). The data
suggests that the critical points measured by decoding and shadow tomography
coincide: $p_{c,<}^{\text{S}}\approx p_{c,<}^{\mathcal{D}}$. Note that both
protocols rely on the same error correction algorithm $\mathcal{E}_E$.

\paragraph{Summary.}

We introduced experimentally accessible protocols to lower and upper bound the
entanglement transition of the projective transverse field Ising model -- which
is not directly accessible due to the postselection problem. Our protocols
combine information retrieval and shadow tomography with sophisticated error
correction algorithms to cope with faulty measurement records.
\cref{fig:shadow}~(c) compares the experimentally accessible quantities from
all protocols with the inaccessible ancilla entanglement entropy. All exhibit a
clear crossing for increasing system size $L=T\to\infty$, indicating the
critical values $p_{c,<}^{\mathcal{D}} \approx p_{c,<}^{\text{S}} \leq p_c \leq
p_{c,>}^{\text{S}}$. Notably, the interval
$\delta:=p_{c,>}^{\text{S}}-p_{c,<}^{\text{S}}$ depends approximately linearly
on the noise rate $\eta$ (inset) and therefore provides an independent
experimental probe of the latter.


\paragraph{Acknowledgments.}

\begin{acknowledgments}
	We acknowledge funding by the  Federal Ministry for Research, Technology and Space (BMFTR) project MUNIQC-ATOMS (Grant No. 13N16070).
\end{acknowledgments}


\paragraph{Data availability.}

The data that support the findings of this article are openly available \cite{data}.

\bibliographystyle{bib/bibstyle}
\bibliography{bib/bibliography}


\setcounter{secnumdepth}{3}

\renewcommand{\thesection}{S\arabic{section}}
\renewcommand{\theequation}{S\arabic{equation}}
\renewcommand{\thefigure}{S\arabic{figure}}
\renewcommand{\thetable}{S\arabic{table}}

\setcounter{section}{0}
\setcounter{equation}{0}
\setcounter{figure}{0}
\setcounter{table}{0}

\clearpage
\section*{Supplementary material}

\section{Error correction on the projective transverse field Ising model}

Here we describe how the projective transverse field Ising model (PTIM) can be
error-corrected using a minimum-weight perfect matching (MWPM) algorithm. In
quantum information, the task of a decoder is to recover the initial state of a
quantum code after noise has affected the system. For information retrieval in
the PTIM, we want to construct a correction operator $C=\prod_iX_i^{c_i}$ from
a correction pattern $\vec{c}=(c_1,\ldots,c_L)$ such that
$\ket{\Psi_0}=C\ket{\Psi_T}$, where $\ket{\Psi_0}$ denotes the initial state
and $\ket{\Psi_T}$ the final state of the random quantum circuit. In the
context of shadow tomography, similar error correcting algorithms are used, but
in this case for classical state prediction.

Given a noisy PTIM, the observer records an \emph{incomplete} measurement
pattern $M^\text{p}=(E^\text{p},S^\text{p})$ and the corresponding measurement
results $M^\text{r}=(E^\text{r},S^\text{r})$ for both types of measurements
$E_i$ and $S_e$.
Due to its incompleteness, the measurement history $M^\text{r}$ corresponds to
an impossible PTIM trajectory in general. For example, consider two consecutive
$E_i$-measurements with different outcomes. This is only possible if the qubit
$i$ was affected by an $S_e$-measurement between the two $E_i$-measurements
which was not recorded. This suggests that inconsistent $E_i$-measurements
contain information on lost $S_e$-measurements. Conversely, the results of
$S_e$-measurements are inconsistent if and only if $E_i$-measurements occurred
that were not recorded.

The idea is to use an error correction algorithm to construct an augmented (and
partially hypothetical) measurement history that it self-consistent and matches
the observed measurement outcomes. The algorithm
$\mathcal{E}_{E,S}(M^{\text{p},\text{r}})$ takes the incomplete measurement
history as input and finds additional $E_i$- and $S_e$-measurements (without
predicting their results) to augment the pattern $M^\text{p}\mapsto\tilde
M^\text{p}$ and make the trajectory consistent. Note that the simplest
algorithm assumes that \emph{every} possible measurement was performed, as this
makes the recorded results trivially consistent. However, this measurement
history is very unlikely, and therefore such a decoder performs sub-optimal in
recovering the initial state. Hence we use a much more sophisticated MWPM
algorithm to make a better prediction for the actual trajectory of the noisy
PTIM, and thereby increase the likelihood of guessing the initial state
correctly.

After the error correction, the augmented measurement pattern
$\tilde{M}^\text{p}$ can not only be used by a decoder $\mathcal{D}$ to
construct the correction pattern $\vec c$ (to recover the initial state
$\ket{\Psi_0}$), but also to predict the final state $\ket{\Psi_T}$ (for a
known initial state $\ket{\Psi_0}$) which can then be used for shadow
tomography (see main text).

In the following subsections, we describe explicitly how the error correction
algorithm $\mathcal{E}_{E,S}$ used in this paper is constructed. The algorithm
is composed of two subroutines: A first correction
$\mathcal{E}_E=\mathcal{E}_E(M^\text{p},S^\text{r})$ uses the recorded
measurement pattern and the results of the $S_e$-measurements to predict
$E_i$-measurements that were not recorded, thereby generating an augmented
pattern $\tilde{E}^\text{p}$.  Subsequently, an algorithm
$\mathcal{E}_S(M^\text{p},E^\text{r})$ utilizes the results of
$E_i$-measurements to generate an augmented pattern $\tilde{S}^\text{p}$ which
makes the $E_i$-measurements consistent. Note that both correction algorithms
do not predict the \emph{outcomes} of the additional measurements. We can then
combine the two augmented measurement patterns to
$\tilde{M}^\text{p}=(\tilde{E}^\text{p},\tilde{S}^\text{p})$. By construction,
this augmented measurement pattern makes all observed results $M^\text{r}$
self-consistent.

\subsection{Error correction algorithm $\mathcal{E}_E$}
\label{app:DecodingErrors}

\begin{figure*}
	\centering
	\includegraphics[width=0.9\linewidth]{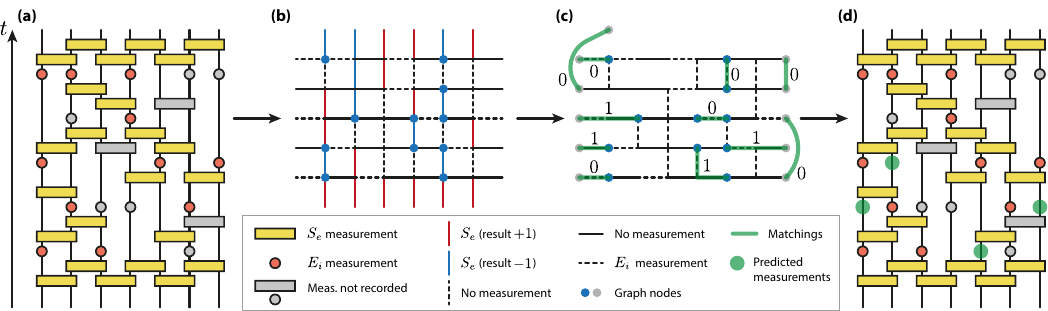}
	\caption{%
		\CaptionMark{MWPM algorithm for the error correction $\mathcal{E}_E$.}
		(a)~Sample PTIM trajectory $M^\text{p}=(E^\text{p},S^\text{p})$ with noise
		(unseen measurements: gray circles and boxes). 
		(b)~The pattern $(E^\text{p},S^\text{p})$ is re-formatted into a grid and
		measurement results $S^\text{r}$ are included (red and blue vertical
		edges). This data is the input of the algorithm $\mathcal{E}_E$. The
		pattern is augmented by nodes at the endpoints of strings of
		$S_e$-measurements with result $-1$ (blue discs).
		(c)~The grid is reduced and converted into a weighted graph by deleting all
		solid vertical edges and adding auxiliary nodes on the boundaries; the edge
		weights count the number of traversed horizontal solid edges in the grid.
		For this graph, a minimum-weight perfect matching is computed (green edges)
		and the matched edges are associated with their corresponding weights.
		Boundary nodes are used to correctly handle the system boundaries.  
		(d)~Matched edges with positive weight correspond to additional
		$E_i$-measurements which augment the original trajectory
		$E^\text{p}\mapsto\tilde E^\text{p}$ (green discs). Note that these
		predictions rarely match the actual missed measurements (cf.\ green and
		gray discs).
	}
	\label{fig:DecodingE}
\end{figure*}

In a first step, we describe the error correction algorithm
$\mathcal{E}_E=\mathcal{E}_E(M^\text{p},S^\text{r})$, which is a modified
version of the MWPM decoder used in Ref.~\cite{Roser_2023}. For the
construction of $\mathcal{E}_E$, one reinterprets a PTIM trajectory in
$1+1$-dimensional ``spacetime'' as a $2$-dimensional error correcting code --
similar to the use of MWPM for decoding two-dimensional surface codes (without
faulty stabilizer measurements) \cite{Duclos-Cianci_2010, Fowler_2012}. 
The correction algorithm is composed of three steps: First, the recorded
measurement pattern $M^\text{p}$ and the results $S^\text{r}$ are used to
construct an abstract weighted graph. Second, a minimum-weight perfect matching
algorithm is used to pair the nodes of this graph. Finally, the matchings are
used to predict an augmented measurement pattern $\tilde{E}^\text{p}$ which
includes additional measurements that make the observed pattern $S^\text{r}$
consistent.  This algorithm is depicted in \cref{fig:DecodingE}:

\cref{fig:DecodingE}~(a) shows the measurement pattern
$M^\text{p}=(E^\text{p},S^\text{p})$ of a sample PTIM trajectory. It includes
an initialization step and a post-processing step where all $S_e$-measurements
are performed. Due to errors, some measurements of both types were not recorded
and are shown in gray.

In a first step, we draw the observed trajectory in a square grid -- only
including the measurements in $M^\text{p}$ which were recorded, see
\cref{fig:DecodingE}~(b). In this notation, the qubits live \emph{between}
vertical lines which represent $S_e$-measurements. The observed results
$S^\text{r}$ are shown as colors of vertical lines. The measurement results
$E^\text{r}$ of $E_i$-measurements are not needed for $\mathcal{E}_E$. The data
in \cref{fig:DecodingE}~(b) represents the \emph{input} for the error
correction algorithm $\mathcal{E}_E$. In a preliminary step, the results
$S^\text{r}$ are used to augment the grid by nodes (blue discs). These are
positioned at the endpoints of vertical blue lines which represent measurement
outcomes $-1$ of $S_e$-measurements (there are no nodes added on the boundary
of the grid).

Next, in \cref{fig:DecodingE}~(c), we modify the grid by removing all
$S_e$-measurements (solid vertical edges) and adding additional nodes on the
left and right boundary of the system at every time step. These additional
nodes are necessary for systems with open boundary conditions in spatial
direction. If (and only if) the total number of nodes (endpoints and boundary)
turns out to be \emph{odd}, an additional boundary node is added outside of the
grid. 
At this stage, one can construct an edge-weighted graph as follows. All nodes
represent nodes of the graph. Edges are created between all nodes which are
connected via the grid (horizontal and vertical, solid and dashed edges). Every
edge is assigned a weight which is a positive integer. The weight corresponds
to the minimal number of horizontal, solid edges on the grid that are traversed
when connecting two nodes. Finally, all boundary nodes are connected to each
other with edges of weight zero. This graph is shown in
\cref{fig:DecodingE}~(c) (ignore the green edges), where the zero-weight
edges connecting boundary nodes are omitted for clarity.

In the next step, one computes a \emph{minimum-weight perfect matching} of this
graph: An MWPM pairs all graph nodes along edges such that the total weight
along matched edges is minimized.
Minimum-weight perfect matchings can be found efficiently using the
\emph{Blossom algorithm} \cite{Edmonds_1965}; here we use an optimized
implementation known as \texttt{Blossom V} by
Kolmogorov~\cite{Kolmogorov_2009}. The resulting minimum-weight perfect
matching is drawn on the graph, including the weights of the matched edges
[green edges in \cref{fig:DecodingE}~(c)]. Note that for a PTIM trajectory
without noise, it is always possible to find a matching of weight zero.

In the last step, a prediction for missing $E_i$-measurements is deduced from
this matching. The matched edges, if drawn on the grid as in
\cref{fig:DecodingE}~(c), traverse dashed vertical and horizontal edges as well
as solid horizontal edges. Every \emph{solid horizontal edge} in the matching
corresponds to a missing $E_i$-measurement. This yields the augmented pattern
$E^\text{p}\mapsto\tilde E^\text{p}$ shown in \cref{fig:DecodingE}~(d).
Importantly, the predicted measurements typically do not coincide with the
real, missed measurements [which are also shown in \cref{fig:DecodingE}~(d) for
comparison]. Instead, the predicted measurements only lead to \emph{one}
possible measurement pattern $\tilde E^\text{p}$ which is consistent with the
recorded measurement results $S^\text{r}$.

It is worthwhile to note that in a similar (classical!) model (where
$E_i$-measurements are replaced by classical bit flips) this error correction
algorithm constructs the \emph{most likely} pattern of additional bit flips
needed to make all parity measurements $S_e$ consistent. However, the PTIM
discussed here contains \emph{projective measurements} $E_i$ instead. The
interpretation of such measurements as ``bit flips'' (indicating whether a
qubit is flipped) is not correct because the actual PTIM states contain
superpositions and entanglement. Nonetheless, this interpretation still
provides some intuition on how (and why) the algorithm $\mathcal{E}_E$ works.
For more details on this point, we refer the reader to Ref.~\cite{Roser_2023}.

When using the error correction algorithm $\mathcal{E}_E$ in the context of
shadow tomography, the PTIM is augmented by an ancilla qubit [recall
\cref{fig:PTIM}\nobreak\space(c) of the main text]. To apply $\mathcal{E}_E$,
one omits the ancilla from the grid in \cref{fig:DecodingE}\nobreak\space(b) to
ensure that only errors on the system qubits $\left\{1,\ldots,L\right\}$ are
predicted. Apart from this, the algorithm remains unchanged.

\subsubsection{Construction of correction operators for information retrieval}

For the decoding protocol described in the main text, the decoder
$\mathcal{D}_E$ uses the previously described algorithm $\mathcal{E}_E$ to
construct a correction operator $C=\prod_iX_i^{c_i}$ with correction pattern
$\vec c=(c_1,\ldots,c_L)\in\left\{0,1\right\}^L$. The final time step that
contains all $S_e$-measurements is assumed to be noise-free and can be used to
reduce the possible correction patterns to only two choices, one of which is
the complement of the other: $C_2=\overline{C}_1:=C_1\prod_{i=1}^LX_i$. 

In a scenario without noise ($\eta=0$), the true correction operator can always
be found deterministically by a classical simulation via the stabilizer
formalism (for a trajectory where the initial cluster survives). An even
simpler method uses an \emph{extended colored cluster model} to track qubit
configurations of clusters (see Ref.~\cite{Lang_2020} and
\cref{subsubsec:ModifiedColoredClusterModel}). This method does not require the
results of $E_i$-measurements and relies only on $S_e$-measurement results to
propagate clusters and their configurations through space and time. For the
same reason, it is not necessary to run the error correction algorithm
$\mathcal{E}_S$ (described in \cref{app:DecodingStabilizers}) for the
construction of a correction pattern, since this algorithm cannot predict
results of predicted $S_e$-measurements. Hence these additional
$S_e$-measurements cannot be used to propagate cluster configurations anyway.

In the presence of noise ($\eta\neq 0$), we proceed as follows. First, using
the algorithm $\mathcal{E}_E$ described above, an augmented pattern
$\tilde{E}^\text{p}$ is constructed. Together with the recorded pattern
$S^\text{p}$, this pattern is then used to perform a classical colored cluster
simulation. 
If this simulation predicts that the initial cluster did \emph{not} survive, no
prediction for the correction operator $C$ can be made; hence we choose $C_1$
or $C_2$ randomly. 
By contrast, if the simulation predicts the survival of the cluster, the
observed results $S^\text{r}$ are used to track the qubit configurations of
clusters through space and time. In this case, the configuration of the final
cluster can be inferred (via the extended colored cluster model) and the
corresponding correction operator is returned. In summary, the algorithm tries
to reproduce the actual PTIM trajectory and uses this information to make a
prediction for a correction pattern $\vec{c}$ and the corresponding correction
operator $C\in\{C_1,C_2\}$.

We note that the same algorithm was already used in Ref.~\cite{Roser_2023}, but
there with a different noise model. Note that for the decoder $\mathcal{D}_E$
the noise that affects $S_e$-measurements plays no role.

\subsubsection{Extended colored cluster model}
\label{subsubsec:ModifiedColoredClusterModel}

Here we briefly review the colored cluster model \cite{Lang_2020} and how it
can be extended to track qubit configurations and predict states in the PTIM.
This (classical) model is a much more efficient alternative to generic
stabilizer simulations of the PTIM.

Consider two qubits in a product state $\ket{++}$. An $S_e$-measurement
entangles the two qubits into a Bell state $\ket{00}+\ket{11}$ or
$\ket{01}+\ket{10}$, depending on the measurement outcome. Crucially, both
states share the same entanglement structure. This mechanism extends to more
qubits: $S_e$-measurements merge Bell pairs to larger, entangled \emph{Bell
clusters}. Conversely, $E_i$-measurements disentangle a single qubit by
removing it from a Bell cluster.

The colored cluster model tracks the entanglement dynamics of the PTIM
efficiently by only modeling Bell clusters. To this end, all qubits are
assigned colors. If qubits share a color, they belong to the same Bell cluster.
(One can consider disentangled qubits as ``single site Bell clusters'' and
assign each their own color.) 
The dynamics of the model follows only two rules:
\begin{enumerate}

	\item If an $E_i$-measurement is performed on a qubit, it is assigned a
		unique new color.

	\item If a $S_e$-measurement is performed between two qubits which have
		different colors, one of the corresponding clusters inherits the color of
		the other cluster. This joins the Bell clusters. If both qubits were
		already part of the same cluster (= had the same color), the measurement is
		deterministic and nothing needs to be done.

\end{enumerate}
To track the survival of encoded amplitudes in the PTIM (i.e., the survival of
an initial Bell cluster) it is useful to initially assign a designated color to
all qubits. If qubits of this particular color are involved in
$S_e$-measurements, the designated color is inherited by the other cluster.
The initial cluster survived if the designated color exists in the final state.

Note that the colored cluster model does not rely on measurement \emph{results}
and thus cannot infer the final \emph{state} of a trajectory. If the PTIM is
initialized in a product state $\ket{\vec{m}}$ and this initial cluster
survives the trajectory, the final state has the form $\ket{\tilde{\vec{m}}}$
with a different qubit configuration $\tilde{\vec{m}}$. A stabilizer simulation
could be used to find this final state. However, an \emph{extended colored
cluster model} can be used to perform the same task much more efficiently. 
To this end, in addition to a color, every qubit is assigned a (classical)
binary state ($0$ or $1$). The initial configuration of these bits is given by
the qubit configuration $\vec{m}$ of the initial state $\ket{\vec{m}}$. 
To evolve these new binary states, another rule extends the colored cluster
model dynamics:
\begin{enumerate}

	\setcounter{enumi}{2}

	\item If an $S_e$-measurement is performed on two qubits, their
		configurations must match the measurement outcome. If they already do,
		nothing needs to be done. If not, all qubits which subsequently inherit a
		new color are also assigned their inverted binary state.

\end{enumerate}
This procedure tracks one of the two possible configurations in every Bell
cluster $\ket{\vec{n}}_A\pm\ket{\overline{\vec{n}}}_A$ and the actual state of
the initial cluster $\ket{\vec m}$. This facilitates an efficient state
prediction if the initial cluster survives.

\subsection{Error correction algorithm $\mathcal{E}_S$}
\label{app:DecodingStabilizers}

\begin{figure*}
	\centering
	\includegraphics[width=0.9\linewidth]{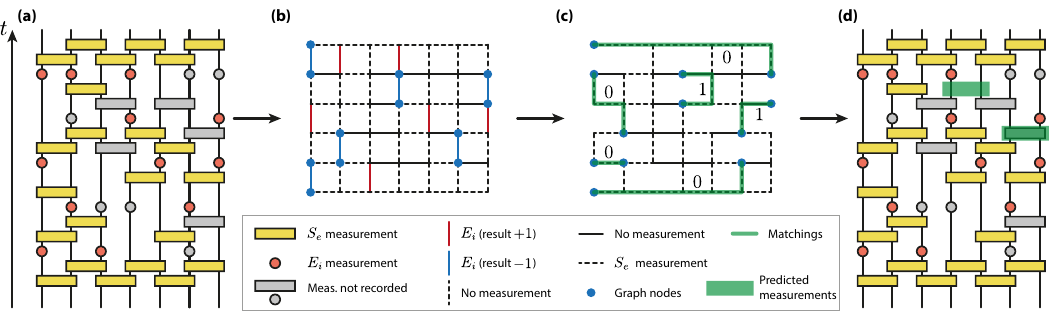}
	\caption{%
		\CaptionMark{MWPM algorithm for the error correction $\mathcal{E}_S$.}
		(a)~Sample PTIM trajectory $M^\text{p}=(E^\text{p},S^\text{p})$ with noise
		(unseen measurements: gray circles and boxes). For demonstration purposes
		we use a slightly different trajectory than in \cref{fig:DecodingE}~(a). 
		(b)~The grid encodes the measurement pattern together with the measurement
		outcomes $E^\text{r}$ (red and blue vertical edges). Note that in
		comparison to \cref{fig:DecodingE}~(b), here solid vertical edges
		correspond to $E_i$-measurements and dashed horizontal edges encode
		$S_e$-measurements (without results). Nodes are assigned to endpoints of
		strings of $E_i$-measurements with outcome $-1$ (blue discs).
		(c)~To construct the reduced grid, all $E_i$-measurements (solid vertical
		edges) are removed. The resulting picture is transformed into a weighted
		graph of which a minimum-weight perfect matching is computed (every solid
		horizontal edge contributes a weight of one). One such matching is drawn on
		the grid (green edges). 
		(d)~Finally, matched edges with non-zero weight are identified with
		additional $S_e$-measurements (green boxes). Note that again the
		predictions typically do not match the actual missed measurements (cf.\
		green and gray boxes).
	}
	\label{fig:DecodingS}
\end{figure*}

The purpose of the error correction algorithm
$\mathcal{E}_S=\mathcal{E}_S(M^\text{p},E^\text{r})$ is to construct an
augmented measurement pattern $S^\text{p}\mapsto\tilde{S}^\text{p}$ which makes
the observed results $E^\text{r}$ consistent. The algorithm is in some sense
``dual'' to the algorithm $\mathcal{E}_E$ in \cref{app:DecodingErrors}. We
first describe the algorithm itself before providing an explanation on the
physics that motivates it below.

We start again with a sample PTIM trajectory shown in \cref{fig:DecodingS}~(a).
As a first step, we restructure the trajectory again into a square grid.
However, in this case, we use the grid \emph{dual} to the one used for the
algorithm $\mathcal{E}_E$ [cf.~\cref{fig:DecodingE}~(b)]. In
\cref{fig:DecodingS}~(b) qubits live on the \emph{vertical edges} and the
pattern $S^\text{p}$ is represented by \emph{horizontal edges}. In particular,
$E_i$-measurements correspond to solid vertical edges and $S_e$-measurements to
dashed horizontal edges. We also encode the results $E^\text{r}$ by colors of
vertical edges. Similar to before, nodes are attached to endpoints of vertical
lines that correspond to $E_i$-measurements with result $-1$.

The subsequent steps are similar to \cref{app:DecodingErrors}, with the
difference that the left and right boundaries of the grid are smooth (no
dangling edges) and no boundary nodes are added. In \cref{fig:DecodingS}~(c) we
assume that the initial state was prepared with positive parity
$\left(\prod_{i=1}^LX_i\right)\ket{\Psi_0}=\ket{\Psi_0}$ (say
$\ket{\Psi_0}\propto\ket{0\ldots0}+\ket{1\ldots1}$). If the parity is negative,
it is necessary to consider ``virtual periodic boundaries'' in time-direction
by adding boundary nodes at every site before the first and after the last time
step (see \cref{subsubsec:extensions} below). 

In the next step [\cref{fig:DecodingS}~(c)] we reduce the grid by removing all
$E_i$-measurements (solid vertical edges). Using the same procedure as for the
correction algorithm $\mathcal{E}_E$, the grid is transformed into an
edge-weighted graph on which a minimum-weight perfect matching is computed.
Here every solid horizontal edge (an $S_e$-measurement missing in $S^\text{p}$)
contributes a weight of one. An MWPM with the corresponding weights is shown in
\cref{fig:DecodingS}~(c). 
Then, matchings with non-zero weight correspond to additional predicted
$S_e$-measurements that lead to an augmented pattern $\tilde{S}^\text{p}$
[shown in \cref{fig:DecodingS}~(d)]. As before, the predictions generally do
not coincide with the actual pattern of non-recorded measurements.

The physical insight from which the error correction algorithm $\mathcal{E}_S$
derives is that the total parity $\prod_{i=1}^{L}X_i$ is a conserved quantity
of the PTIM dynamics (as it commutes with all measurements). 
First, consider a noise-free PTIM trajectory. Every $E_i$-measurement creates a
single-site Bell cluster $\ket{0}_i\pm\ket{1}_i$ with parity (= relative sign)
$\pm 1$ fixed by the measurement outcome.  Whenever two Bell clusters merge via
an $S_e$-measurement, the product of their parities determines the parity of
the new combined cluster (independent of the result of the $S_e$-measurement).
Now consider the situation where $E_i$-measurements are performed on a number
of qubits which are subsequently merged into a Bell cluster (by
$S_e$-measurements). The product of the outcomes of these $E_i$-measurements
determines the parity of the Bell cluster. If the cluster is later destroyed by
other $E_i$-measurements, the product of their outcomes corresponds to a
measurement of the parity of the Bell cluster. One can convince oneself that
this is also true for more complicated cluster histories: When drawing a Bell
cluster which eventually dies out in spacetime, it can be surrounded by a loop
of $E_i$-measurements and missing $S_e$-measurements. The product of outcomes
of these $E_i$-measurements is necessarily $+1$. Therefore missing
$S_e$-measurements in a \emph{noisy} PTIM can be detected by identifying
spacetime loops of $E_i$-measurements which multiply to $-1$. This situation
occurs when two clusters were actually connected by an $S_e$-measurement that
was not recorded. The correction algorithm $\mathcal{E}_S$ described above
finds a minimal extension $S^\text{p}\mapsto\tilde{S}^\text{p}$ of the pattern
of $S_e$-measurements which joins clusters in spacetime such that all loops
around clusters have parity $+1$ and are therefore consistent.

\subsubsection{Extensions needed for the shadow tomography protocol}
\label{subsubsec:extensions}

In the main text, we employ the error correction algorithm $\mathcal{E}_S$ to
upper bound the entanglement entropy of an ancilla qubit via shadow tomography
[recall \cref{fig:PTIM}~(c)]. This requires two extensions to the procedure
described above. 

First, to accommodate the entanglement transfer at the end, an additional time
step is added to the grid, including the (noise-free) $E_i$-measurements on the
qubits $\left\{1,\ldots,L-1\right\}$. Second, as the initial Bell cluster
includes the ancilla qubit, the parity of the subsystem which contains only
system qubits is initially not well-defined. Therefore paths of
$E_i$-measurements and missing $S_e$-measurements cannot be closed along the
lower and upper boundary of the grid ($t=0$ and $t=T$). To account for the
possibility that the parity of $E_i$-measurements along such paths can multiply
to $\pm 1$, we add artificial boundary nodes before the first and after the
last time step. This creates ``virtual'' periodic boundaries on the grid in
time-direction and ensures that in the noise-free case a zero-weight perfect
matching always exists.

These modifications are not shown in \cref{fig:DecodingS}~(b) for the sake of
clarity. Note that the ancilla qubit is assumed to be noise-free so that no
measurements affect this qubit. We enforce this by only including the system
qubits $\left\{1,\ldots,L\right\}$ in the grid.

\section{Classical state prediction with error correction}
\label{app:ClassicalStatePredictionViaDecoding}

In the shadow tomography protocol described in the main text, the error
correction algorithms $\mathcal{E}_{E}$ and $\mathcal{E}_S$ are used for the
classical prediction of the final reduced density matrix
$\rho_{M^{\text{p},\text{r}},\left\{\text{a},L\right\}}^\text{C}$ of the
ancilla qubit and the last system qubit $L$. 
After the entanglement transfer step [see \cref{fig:PTIM}~(c)], the two qubits
are in one of eight possible states. (We describe these reduced density
matrices in terms of the generators of their stabilizer group.) If the initial
cluster survived, the ancilla qubit is still fully entangled with the system
qubit and the final state is stabilized by $\langle\pm Z_\text{a}Z_L,\pm
X_\text{a}X_L\rangle$. By contrast, if the initial cluster was destroyed on the
system qubits, the final state is a product state of two single-qubit states
and stabilized by $\langle \pm X_\text{a},\pm X_\text{a}X_L\rangle$. 
We now describe a classical algorithm which takes as input an observed PTIM
trajectory $M^{\text{p},\text{r}}$ and produces a prediction for the final
state of the two qubits (that then can be used for shadow tomography).

As every final state is stabilized by $\pm X_\text{a}X_L$, our first goal is
to determine the sign of this generator. Since the PTIM conserves the total
parity $X_\text{a}X_1\cdots X_L$, this step is trivial: In the entanglement
transfer step, the partial parity $X_1\cdots X_{L-1}$ is measured. The result
of this measurement can then be multiplied by the (known) parity of the initial
state to determine the sign of the generator $\pm X_\text{a}X_L$. 

In the next step, the algorithm decides whether or not the two qubits are
entangled. To this end, the two algorithms $\mathcal{E}_{E}$
(\cref{app:DecodingErrors}) and $\mathcal{E}_S$
(\cref{app:DecodingStabilizers}) are employed to predict a consistent
measurement pattern
$\tilde{M}^\text{p}=(\tilde{E}^\text{p},\tilde{S}^\text{p})$.  Subsequently, a
colored cluster simulation is performed on this pattern
(\cref{subsubsec:ModifiedColoredClusterModel}). If the initial Bell cluster is
found to have survived on the system qubits, the algorithm concludes that the
ancilla qubit must be entangled with the system; otherwise the two qubits are
assumed to be in a product state. 
To determine the sign of the remaining generator, the algorithm proceeds
differently for the two cases:
\begin{itemize}

	\item 
		If an \emph{entangled} state is predicted, the algorithm must determine the
		sign of the generator $\pm Z_\text{a}Z_L$. 
		In the noise-free case, this can be done by tracing the initial cluster
		through spacetime via an extended colored cluster simulation. 
		If there is noise, only recorded $S_e$-measurements (with known result) can
		be used to predict the sign -- the additional measurements predicted by
		the error correction subroutine $\mathcal{E}_S$ are useless for this
		purpose because their outcome is unknown. 
		To check whether the sign can be predicted, a second extended colored
		cluster simulation is performed, but now on the pattern
		$(\tilde{E}^\text{p},S^\text{p})$ which \emph{ignores} the additional
		$S_e$-measurements predicted by $\mathcal{E}_S$.
		If the cluster survives according to this simulation, the sign of the
		generator $\pm Z_\text{a}Z_L$ can be inferred from the \emph{observed}
		$S_e$-measurement outcomes. By contrast, if the cluster is now destroyed,
		it survived under the augmented pattern $\tilde M^\text{p}=(\tilde
		E^\text{p},\tilde S^\text{p})$ only due to additional $S_e$-measurements
		predicted by $\mathcal{E}_S$. In this case, the sign of the generator
		cannot be inferred and the algorithm returns the partially mixed state
		$\langle\pm X_\text{a}X_L\rangle$.

	\item If a \emph{product} state is predicted, the algorithm must determine
		the sign of the generator $\pm X_\text{a}$. 
		In a noise-free PTIM, the initial Bell cluster was destroyed on the system
		qubits by $E_i$-measurements on all qubits. This can be interpreted as a
		percolation in spacetime from the leftmost to the rightmost qubit.
		[Consider the grid shown in \cref{fig:DecodingE}~(b). Were the initial
		cluster to be destroyed (which is not the case for the shown trajectory), a
		horizontal path along only dashed edges would exist which includes all
		qubits.] The product of all $E_i$-measurement outcomes along such a
		percolation path is the parity measured on the system. One can then multiply
		this parity with the (known) parity of the initial state to infer the
		missing parity of the ancilla qubit and decide on the sign of the generator
		$\pm X_\text{a}$. 
		In a \emph{noisy} PTIM, this construction is only possible if all
		$E_i$-measurement results along a percolation path that connects the two
		spatial boundaries are recorded.  To check this, we perform a colored
		cluster simulation on the pattern
		$\left(E^\text{p},\tilde{S}^\text{p}\right)$ which excludes the additional
		$E_i$-measurements predicted by $\mathcal{E}_E$. If the cluster survives
		according to this simulation, it is only destroyed due to additional
		$E_i$-measurements predicted by $\mathcal{E}_E$. In this case, the
		algorithm cannot infer the sign of the generator and returns the partially
		mixed state $\langle\pm X_\text{a}X_L\rangle$. By contrast, if the
		cluster still does not survive, a percolation path can be found on which
		all $E_i$-measurement outcomes are known so that its parity can be used to
		infer the sign of the generator $\pm X_\text{a}$.

\end{itemize}

\begin{figure}[tbp]
	\centering
	\includegraphics[width=0.9\linewidth]{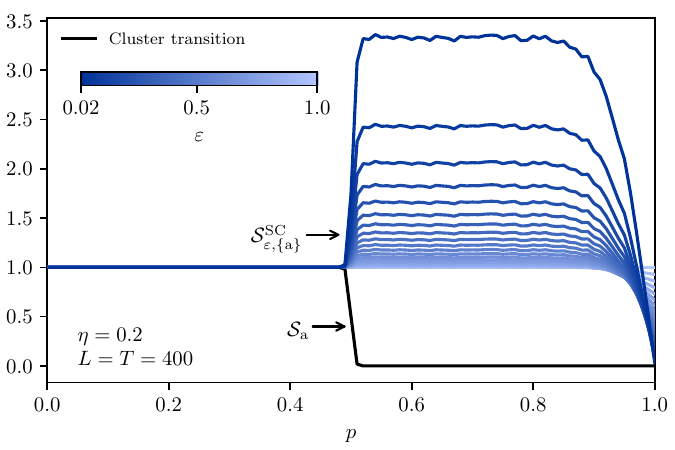}
	\caption{%
		\CaptionMark{Shadow tomography of a noisy PTIM without error correction.}
		Upper bound for the cluster transition obtained via shadow tomography based
		on a na\"ive state prediction that does \emph{not} use an error correction
		algorithm. Clearly this procedure yields no useful upper bound. The shown
		data is based on $10^5$ samples.
	}
	\label{fig:shadow_nodec}
\end{figure}

In conclusion, this algorithm returns a stabilizer generator which can then be
transformed into a density matrix
$\rho_{M^{\text{p},\text{r}},\left\{\text{a},L\right\}}^\text{C}$. This
prediction can then be used to find the lower bound on the entanglement transition
in \cref{fig:shadow} (b). 
Note that the state prediction for the reduced density matrix
$\rho_{M^{\text{p},\text{r}},\left\{\text{a}\right\}}^\text{C}$ of the ancilla
qubit \emph{alone} is needed for both the lower and the upper bound. It is
defined and can easily be computed via the partial trace:
\begin{align}
	\rho_{M^{\text{p},\text{r}},\left\{\text{a}\right\}}^\text{C}
	:=\operatorname{Tr}_L\left[
		\rho_{M^{\text{p},\text{r}},\left\{\text{a},L\right\}}^\text{C}
	\right]
	\,.
\end{align}

\section{Shadow tomography without error correction}
\label{app:ShadowTomographyWithoutDecoding}

As a sanity check, we demonstrate that the non-trivial upper (and lower) bounds
obtained via shadow tomography (see \cref{fig:shadow} of the main text)
crucially rely on the performance of the error correction algorithm
$\mathcal{E}_{E,S}$. To do so, we show that state prediction without an error
correction subroutine leads to trivial upper bounds for the entanglement
transition in the presence of noise.

We start with the description of an algorithm for the prediction of the state
$\rho_{M^{\text{p},\text{r}},\left\{\text{a}\right\}}^\text{C}$ of the ancilla
qubit which does \emph{not} use an error correction subroutine.
As a first step, a colored cluster simulation is performed based on the
observed pattern $M^\text{p}$ (\cref{subsubsec:ModifiedColoredClusterModel}).
If this simulation suggests that the initial Bell cluster survived on the
system qubits, the prediction for the ancilla qubit must be a fully mixed
state:
$\rho_{M^{\text{p},\text{r}},\left\{\text{a}\right\}}^\text{C}=\tfrac{1}{2}\mathds{1}$.
Conversely, if the simulation suggests that the Bell cluster was destroyed, the
ancilla qubit is predicted to be in a pure state $\ket{\pm}$ polarized in
$x$-direction. As described in \cref{app:ClassicalStatePredictionViaDecoding},
in this case there exists a path of $E_i$-measurements and omitted
$S_e$-measurements which traverses spacetime from qubit $1$ to $L$. As a first
step, this percolation path is identified. To determine the pure state
($\ket{+}$ or $\ket{-}$) of the ancilla qubit, the parity of the identified
path (computed from the involved $E_i$-measurements) is multiplied with the
parity of the initial state.
In contrast to the error correction-based approach in
\cref{app:ClassicalStatePredictionViaDecoding}, this state prediction is only
based on the \emph{observed} pattern $S^\text{p}$ -- instead of the augmented
pattern $\tilde{S}^\text{p}$ obtained by error correction (which would identify
some of the percolation paths as unreliable).

We show simulations for this error correction-free shadow tomography in
\cref{fig:shadow_nodec} for the noise rate $\eta=0.2$ and system size
$L=T=400$. We verified for different system sizes that the upper bound on the
entanglement transition approaches $p\to 1$ for $L\to\infty$. In particular --
and in contrast to our error correction-based results in \cref{fig:shadow}~(c) -- no
crossing appears. This obviously renders shadow tomography useless to measure
an upper bound on the entanglement transition of the PTIM in the presence of
noise. 

We conclude that to obtain a meaningful and noise-resilient upper (and lower)
bound via shadow tomography, it is mandatory to employ performant error
correction algorithms (like MWPM) to counter experimental artifacts.

\end{document}